# Hierarchical Framework for Predicting Entropies in Bottom-Up Coarse-Grained Models

Jaehyeok Jin[*] and David R. Reichman[*]

*3000 Broadway, Department of Chemistry, Columbia University, New York, NY 10027, USA*

* Corresponding author: jj3296@columbia.edu, drr2103@columbia.edu

**Abstract**

The thermodynamic entropy of coarse-grained (CG) models stands as one of the most important properties for quantifying the missing information during the CG process and for establishing transferable (or extendible) CG interactions. However, performing additional CG simulations on top of model construction often leads to significant additional computational overhead. In this work, we propose a simple hierarchical framework for predicting the thermodynamic entropies of various molecular CG systems. Our approach employs a decomposition of the CG interactions, enabling the estimation of the CG partition function and thermodynamic properties *a priori*. Starting from the ideal gas description, we leverage classical perturbation theory to systematically incorporate simple yet essential interactions, ranging from the hard sphere model to the generalized van der Waals model. Additionally, we propose an alternative approach based on multiparticle correlation functions, allowing for systematic improvements through higher-order correlations. Numerical applications to molecular liquids validate the high fidelity of our approach, and our computational protocols demonstrate that a reduced model with simple energetics can reasonably estimate the thermodynamic entropy of CG models without performing any CG simulations. Overall, our findings present a systematic framework for estimating not only the entropy but also other thermodynamic properties of CG models, relying solely on information from the reference system.

**TOC Graphic**

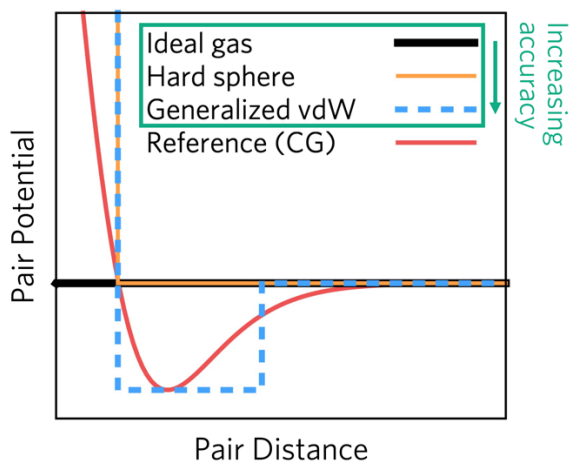

## 1. Introduction

In order to gain insights into chemical and physical processes at an atomistic level, computer simulations have become an essential tool in the fields of chemistry, physics, and biology. Conventionally, this entails performing molecular dynamics (MD) or Monte Carlo (MC) simulations[1-7] using effective Hamiltonians designed to capture a range of interactions at the atomistic level, i.e., non-bonded, pair-bonded, angular, and dihedral interactions.[8-13] However, such a description at the fully atomistic level is often computationally expensive to simulate for complex systems over long length and time scales.[14, 15] To surmount this limitation, coarse-grained (CG) modeling has emerged as a strategy to reduce the number of complex atomistic degrees of freedom, yielding a simpler system that can be simulated for effectively long times.[15-25] Thus, eliminating unnecessary degrees of freedom can greatly enhance the spatiotemporal scales explored in computational simulations.[26-32]

Since simplified CG systems should exhibit behaviors that are as close as possible to their corresponding fine-grained (FG) counterparts, bottom-up CG approaches have been designed to derive CG interactions based on FG statistics.[15, 17-19, 21, 22, 24, 33, 34] These bottom-up approaches are rooted in statistical mechanical methodologies designed to recapitulate important structural correlations[35, 36] rather than relying on an empirical determination from top-down approaches.[34, 37-40] In pursuit of designing bottom-up CG models, however, we note that the thermodynamic properties of CG systems deviate from those of the FG systems in practical scenarios—an inconsistency commonly known as the representability problem.[21] In principle, the "exact" CG potential, determined by accurately integrating over the FG degrees of freedom, should be able to faithfully reproduce the FG free energy and thermodynamic properties through appropriate derivatives of the free energy. However, due to the approximate nature when determining the many-dimensional CG interactions and their dependence on thermodynamic variables, inconsistencies between the FG and the CG properties often arise. To address this issue, numerous attempts have been made.[41-44]

Among various thermodynamic properties, our particular focus in this paper is on entropy. In CG modeling, the entropy holds privileged importance as it directly accounts for the reduced information resulting from the coarse-graining process. For example, relative entropy minimization formally demonstrates that the entropy is the target object to be optimized when deriving CG models.[36, 45-47] Additionally, given that bottom-up CG interactions are essentially a many-body potential of mean force (PMF), which is a free energy quantity,[48, 49] the difference between the FG and CG entropies amounts to the temperature dependence of the effective CG potentials. Therefore, a correct understanding of the so-called "missing entropy" is pivotal for achieving transferability of CG interactions.[21, 24, 41, 50, 51] While steady effort has been directed toward elucidating the physical nature of this missing entropy through the development of theoretical and computational methodologies,[42, 43, 52-54] relatively little attention has been paid to the CG entropy itself. Due to the non-trivial and many-body nature of CG interactions, calculating the CG entropy requires not only parametrizing the CG interactions but also conducting CG simulations and subsequently employing existing free energy methods to determine the entropy in a post-processing step. Such an extensive undertaking may appear contradictory to the reductionist philosophy inherent in CG modeling.

Motivated by the need to overcome this computational overhead for arbitrary CG systems, our primary goal in this paper is to develop a statistical mechanical framework capable of estimating the correct thermodynamic entropy of the CG system without necessitating CG simulations. In other words, we aim to develop a predictive theory that can reasonably estimate the entropy of complex CG molecules based solely on their model characteristics. This objective may seem challenging given the complexity of bottom-up CG models; however, this difficulty can be overcome. In particular, despite the intricate form of the parametrized CG interaction, if one could effectively decompose it into separate and analytically tractable CG interactions, additional CG simulations would become unnecessary. Since this partitioning is highly non-trivial and varies across different systems, we attempt to approach this problem in a systematic manner inspired by the classical perturbation theory of liquids. As initially proposed by Zwanzig[55] and elaborated later by Barker and Henderson,[56, 57] as well as Weeks, Chandler, and Andersen,[58-60] the effective structure of liquids is primarily determined by their short-range repulsions, often conceived as "hard sphere" repulsions. In the perturbation approach, the hard sphere repulsions serve as the main reference interaction, while less significant long-range interactions (usually attractions) are treated as perturbative terms. Early efforts in computer simulations of simple, analytical liquid models successfully calculated thermodynamic and dynamic properties based on the hard sphere contributions,[61, 62] with further improvements achieved by carefully incorporating perturbative terms. However, it is less clear if this perturbative approach can be faithfully applied to molecular CG models.

Our starting reference is the ideal gas system, where particle interactions are absent. The entropy of an ideal gas is exactly determined from simple properties (number density and molecular weight), providing a basic estimate of the expected entropy under specific conditions. This assessment helps gauge the practicality of using the ideal gas approximation to estimate the entropy of non-ideal CG liquids. Yet, the ideal gas description has a fundamental limitation in describing liquid phases, given an ideal gas occupies no volume. In order to address the concept of free volume in condensed phases,[63-65] we estimate the effective free volume by considering the repulsive contribution from hard spheres, allowing for the incorporation of the non-trivial volumes of CG liquids. While a hard sphere description may seem an extreme one for modeling CG systems with complex many-body interactions, we are particularly motivated by the recent findings indicating that a hard sphere treatment of bottom-up CG systems offers an efficient and accurate way to determine dynamical properties.[66-68] Building upon this research, our work extends beyond investigating the dynamics of CG systems to the determination of the thermodynamic properties.

Despite the effectiveness of hard sphere models in reproducing structural and dynamic properties, they may yield inaccurate thermodynamic properties due to the absence of attractive perturbations. For example, if we only consider repulsions and neglect attractive forces, internal energies will consistently appear positive, whereas in realistic systems with non-negligible attractions, they can be negative. In this light, hard sphere models may not be the most suitable choice for evaluating thermodynamic properties. Therefore, we assess the performance and necessity of addressing these long-range attractions within the hard sphere description. As we will discuss, incorporating long-range attractions is considerably more complex than addressing short-range repulsions. As CG interactions in liquids often exhibit relatively simple interaction profiles with a few attractive or repulsive wells at long ranges, it becomes reasonable to approximate these contributions as simplified well-like interactions. Therefore, as part of our predictive framework, we will incorporate attractive contributions into the van der Waals equation.[69] In the generalized van der Waals theory,[70-77] partition functions and related thermodynamic properties can be analytically derived, similar to the hard sphere case, assuming relatively simple attractive interactions. Compared to simple and commonplace atomistic force fields, such as those in molecular mechanics,[8-13] the bottom-up nature of CG interactions makes this mapping less straightforward, given that CG interactions become effective free energy quantities and vary with system conditions. Hence, we will leverage our understanding of the temperature dependent nature of CG interactions to correctly capture the renormalized FG partition function. This approach aims to accurately approximate the CG entropy by parameterizing over different thermodynamic state points.

Our hierarchical approach, ranging from the ideal gas to the hard sphere and the generalized van der Waals models, is designed to predict the CG entropy of molecular liquids. However, approximating these simplified models often requires knowledge of additional thermodynamic parameters in advance to estimate the thermodynamic properties from the FG statistics. Alternatively, a different theoretical approach, based on the work by Green[78] and Wallace,[79-82] can overcome this limitation by examining the residual entropy (sometimes referred to as "excess entropy" in the literature). The residual entropy is the difference between the system's entropy and that of an ideal gas under the same conditions. Baranyai and Evans further pursued this direction, expressing the residual entropy directly as an ensemble average of many-body functions using the multiparticle correlation expansion.[83] Since the multiparticle expansion approach avoids the need for strong model approximations, it is of great interest to explore whether this approach can further refine the generalized van der Waals treatment.

Altogether, our objective in this paper is to assess the accuracy of each step in our systematic approach by comparing the estimated entropy with the reference ("actual") CG entropy obtained from the CG simulations. Through this process, we aim to identify the most efficient and accurate method for evaluating the thermodynamic properties of CG models *a priori*. The remainder of this paper is organized as follows: Section 2 discusses how to map the reference CG system to the approximate models and how to evaluate the CG entropy within these models. Subsequently, in Sec. 3, we will apply this framework to the cases of liquid methanol and chloroform, respectively.

## 2. THEORY
### 2.1. Estimating the CG Entropy: Systematic Approach
In this section, we will demonstrate the systematic nature of the proposed approach by deriving an expression for the CG entropy. Our focus here is on the single-site CG resolution, which establishes a physically intuitive correspondence between entropy in the CG and FG systems.

The total entropy in the CG system $S_{CG}$ can be exactly expressed as

$$S_{CG} = S_{trans}^{CG} = -k_B \int \int d\mathbf{P}^N d\mathbf{R}^N p(\mathbf{P}^N, \mathbf{R}^N) \ln[(2\pi\hbar)^{3N} p(\mathbf{P}^N, \mathbf{R}^N)],$$

(1)

where $p(\mathbf{P}^N, \mathbf{R}^N)$ is the probability distribution function for the CG phase space variables. We also note that the Gibbs definition of the entropy $[-k_B \int d\mathbf{P} d\mathbf{R} \, p(\mathbf{P}, \mathbf{R}) \ln p(\mathbf{P}, \mathbf{R})]$ does not explicitly contain the $\ln[(2\pi\hbar)^{3N}]$ term, but one should include this factor for the quantum and indistinguishability corrections when integrating over phase space.[84] However, since this correction term is constant, the remaining sections will primarily consider the Gibbs definition of entropy. Since $\mathbf{P}^N$ and $\mathbf{R}^N$ are not coupled in our CG simulations of interest, these two terms can be separated as

$$S_{\text{trans}}^{\text{CG}} = -k_B \langle \ln[\exp(-\beta T_{\text{CG}}(\mathbf{P}^N))] \rangle - k_B \langle \ln[\exp(-\beta U_{\text{CG}}(\mathbf{R}^N))] \rangle + k_B \ln Q_{\text{CG}}, \quad (2)$$

where $T_{\text{CG}}(\mathbf{P}^N) = \sum_I^N \mathbf{P}_I^2 / 2M_I$, $U_{\text{CG}}(\mathbf{R}^N)$ is the CG configurational energy, and $Q_{\text{CG}}$ is the full CG partition function. Even though the momentum contribution can be separated from the configurational contributions and expressed analytically as

$$-k_B \int d\mathbf{P}^N d\mathbf{R}^N p(\mathbf{P}^N, \mathbf{R}^N) \ln[\exp(-\beta T_{\text{CG}}(\mathbf{P}^N))] = \frac{1}{T} \langle \sum_I^N \frac{\mathbf{P}_I^2}{2M_I} \rangle = \frac{3}{2} N k_B, \quad (3)$$

the analytical form for the configurational part $-k_B \langle \ln[\exp(-\beta U_{\text{CG}}(\mathbf{R}^N))] \rangle$ is practically impossible to determine due to the non-trivial nature of the CG interactions $U_{\text{CG}}(\mathbf{R}^N)$. Our proposed framework approaches this problem by dividing $U_{\text{CG}}(\mathbf{R}^N)$ into a set of simple potentials using a perturbative argument; in the simplest case, where there is no interaction, we can derive the entropy term by sequentially including the strongest interaction and other small perturbations in a systematic way to account for the full $S_{\text{trans}}^{\text{CG}}$ contribution.

Equation (2) suggests that this perturbative treatment will result in additive corrections to the entropy. We decompose $U_{\text{CG}}(\mathbf{R}^N)$ into the following sub-interactions:

$$U_{\text{CG}}(\mathbf{R}^N) \approx 0 + U_{\text{hs}}(\mathbf{R}^N) + U_{\text{g-vdW}}(\mathbf{R}^N), \quad (4)$$

where 0 corresponds to the ideal gas interaction, $U_{\text{hs}}(\mathbf{R}^N)$ denotes the hard sphere interaction, and $U_{\text{g-vdW}}(\mathbf{R}^N)$ is the generalized van der Waals potential. Since Eq. (2) has a $\ln[\exp(-\beta U_{\text{CG}}(\mathbf{R}^N))]$ term, the perturbative decomposition given in Eq. (4) yields

$$-k_B \langle \ln[\exp(-\beta U_{\text{CG}}(\mathbf{R}^N))] \rangle = \frac{1}{T}[\langle 0 \rangle + \langle U_{\text{hs}}(\mathbf{R}^N) \rangle + \langle U_{\text{g-vdW}}(\mathbf{R}^N) \rangle], \quad (5)$$

and the free energy contribution can also be decomposed according to the classical perturbation argument. We will later examine the decomposition of the free energy in more detail. Altogether, the separability of Eq. (5) indicates that the CG entropy can be estimated in a systematic manner.

## 2.2. Ideal Gas: Sackur-Tetrode Equation

As discussed earlier in the Introduction, the simplest approximation one can make for the CG system is the ideal gas approximation, where there are no interactions between the CG particles. In the case of a single-site CG system, the ideal gas approximation leads to a CG entropy following the Sackur-Tetrode equation,[85, 86]

$$S_{\text{trn}}^{(\text{id})} = N k_B \left[ \ln \frac{V}{N} \left( \frac{2\pi m k_B T}{h^2} \right)^{\frac{3}{2}} + \frac{5}{2} \right], \quad (6)$$

where $N$ denotes the number of the CG particles at volume $V$ and temperature $T$, with the constants $k_B$ (Boltzmann) and $h$ (Planck). Equation (6) describes all possible microstates that the CG particle can explore under a given system condition. This approach completely ignores the effective CG interactions. Therefore, the configurational integral $\mathbb{Z}_{\text{CG}}$, defined as $\int d\mathbf{R}^N \exp[-\beta U_{CG}(\mathbf{R}^N)]$, reduces to the volume term ($V^N$), and the ideal gas partition function is expressed as

$$Q_{CG}^{(id)} = \frac{1}{N!}\left(\frac{V}{\Lambda^3}\right)^N,$$

(7)

where $\Lambda$ denotes the thermal de Broglie length, $\sqrt{h^2/2\pi m k_B T}$, which includes only the momentum contribution from the ideal gas description. In Sec. 3, we will assess the reliability of this crude approximation for estimating entropy.

### 2.3. Estimating the CG Entropy: Mapping to Hard Sphere

A realistic model for predicting the CG entropy should account for the non-ideal gas nature of CG systems. Based on our recent findings, one possible direction would be to interpret CG particles as hard spheres. In this context, attractive interactions, which are typically more long-ranged than the repulsive ones, are effectively canceled out due to the uniformly distributed molecules in dense conditions. As initially proposed by Widom[87] and subsequently developed by Barker and Henderson[56, 57] as well as Weeks, Chandler, and Andersen,[58-60] a hard sphere description for dense fluids serves as a powerful approximation for understanding various static and dynamic properties of dense liquids.[88-91]

In this section, we will briefly review how to faithfully capture the hard sphere nature of CG liquids. Since hard sphere liquids can be defined solely based on their packing fraction $\eta$, our primary objective is to determine $\eta$ in a way that accurately recapitulates the structural and thermodynamic characteristics of the reference CG model. The simplest way to determine $\eta$ is to find the effective hard sphere diameter (or radius) of a CG system from the definition of the packing fraction

$$\eta = \frac{\pi}{6}\sigma_{EHSD}^3 \rho_b,$$

(8)

where $\sigma_{EHSD}$ is the effective hard sphere diameter, and $\rho_b$ is the bulk number density. Introducing an additional "hard sphere layer" to CG models allows us to simplify the many-body nature of CG systems into a straightforward hard sphere description, and thus various complex CG properties can be expressed as analytic functions within the hard sphere framework.[67]

One of the most straightforward approaches to determining the effective hard sphere diameter is based on the interaction profile of the CG models. Hard spheres exhibit infinite repulsions when the pair distance is shorter than $\sigma_{EHSD}$. Therefore, a physically sound method is to extract the "short-range repulsive interactions" from CG models that have finite repulsive interactions. Barker and Henderson showed that, within their perturbation theory, this amounts to the vanishing of the first-order correction in the free energy term of hard spheres.[56, 57] This method can be extended to CG systems, where the effective hard sphere diameter $\sigma_{BH}$ can be defined using the effective CG potential $U(R)$

$$\sigma_{BH} = \int_0^{R_0} [1 - \exp(-\beta U(R))] \cdot dR,$$

(9)

where $R_0$ is the minimum distance at which the CG interactions are no longer repulsive [$U(R_0) = 0$].

Equation (9) can also be understood as the effective repulsion weighted by its Boltzmann factor. Unlike simple analytical potentials, our CG model approximates the many-body CG PMF as a pair potential, denoted as $U(R)$ in Eq. (9). Due to the limited expressiveness of pairwise basis sets, this approach necessitates different CG pair potentials at various state points. Consequently, in order to represent the hard sphere nature in the reference atomistic system, we would like to emphasize that a thermodynamically consistent approach should estimate a temperature dependent hard sphere diameter from different CG pair potentials. A similar idea has been applied in describing CG dynamics, where temperature dependent CG potentials were utilized to estimate effective hard sphere diameters for an accurate recapitulation of dynamical properties.[66-68] We also note that extracting the temperature dependent nature of interaction model parameters should be applied throughout this section for different hierarchies of descriptions. In this context, the difference in entropy between the FG model and its CG counterpart should correspond to the FG degrees of freedom eliminated by the CG mapping, as well as the entropic effects arising from the temperature dependence of $U(R)$. Further analyses investigating this aspect can be found in Refs. 43 and 53.

### 2.4. Estimating the CG Entropy: Hard Sphere Entropy

There are two main advantages to approximating CG systems as hard spheres. Firstly, unlike the ideal gas approximation, the CG energetics (and other properties) are no longer zero, which makes this approach more realistic. Furthermore, due to the relatively simple nature of hard spheres, their thermodynamic properties can be expressed in an analytical form.[88-91]

*2.4.1. Residual Entropy of Hard Spheres.* To evaluate the thermodynamic properties from the CG partition function introduced in Eq. (7), a common approach in perturbation theory is to separate the ideal gas contribution $Q_{id}$ from the non-ideal contribution $Q_{ex}$

$$Q_{CG} = \left(\frac{1}{N!}\left(\frac{V}{\Lambda^3}\right)^N\right) \cdot \left(\frac{1}{V^N}\int d\mathbf{R}^N \exp[-\beta U_{CG}(\mathbf{R}^N)]\right) \coloneqq Q_{id}Q_{ex}(N,V,T). \quad (10)$$

Then, the CG free energy $W_{CG}$ can be divided into ideal and non-ideal contributions:
$$W_{CG} = -k_B T \ln Q_{CG} = -k_B T \ln Q_{id} - k_B T \ln Q_{ex} = W_{id} + W_{ex}. \quad (11)$$
Note that this decomposition of $W_{CG}$ demonstrates the consistency of our approach with Eq. (5) in Sec. 2.1.

While the ideal gas contribution to the thermodynamic properties can be exactly described, the residual thermodynamic properties of hard spheres are often approximated using equations of state (EOSs). EOSs are simple relationships that describe pressure, volume, and temperature and are often expressed as a function of the compressibility factor
$$Z = \frac{P}{\rho k_B T}. \quad (12)$$
Among the numerous empirical EOSs proposed in the literature,[90, 92-97] we specifically choose to employ the Carnahan-Starling EOS,[98] denoted here as $Z_{CS}$, because it provides an accurate description over a wide range of packing densities, spanning from stable to metastable regimes, with a relatively simple functional form[99, 100]
$$Z_{CS} = \frac{1 + \eta + \eta^2 - \eta^3}{(1-\eta)^3}.$$

(13)

For a more detailed discussion of the EOS choices for various CG systems, interested readers may refer to Ref. 67. Using the Carnahan-Starling EOS, the residual thermodynamic properties can be calculated as follows. Since the thermodynamic properties $X$ of interest primarily depend on $V$ and $T$, the residual property can be calculated as

$$X_{\text{ex}}(V,T) = \int_{\infty}^{V} \left[\left(\frac{\partial X}{\partial V}\right)_T - \left(\frac{\partial X}{\partial V}\right)_T^{\text{id}}\right] dV.$$

(14)

In order to compute the residual entropy from the hard sphere contribution, one needs to determine $W_{\text{ex}}(V,T)$ and $U_{\text{ex}}(V,T)$ since $TS_{\text{ex}}(V,T) = U_{\text{ex}}(V,T) - W_{\text{ex}}(V,T)$. These thermodynamic properties can be obtained from the residual pressure, which is calculated by applying Eqs. (13) to (14), resulting in

$$\frac{P_{\text{ex}}(V,T)}{\rho k_B T} = \frac{4\eta - 2\eta^2}{(1-\eta)^3}.$$

(15)

Using Eq. (14), the residual Helmholtz free energy and internal energy functionals are derived as

$$\frac{W_{\text{ex}}(V,T)}{RT} = \frac{4\eta - 3\eta^2}{(1-\eta)^2},$$

(16a)

$$U_{\text{ex}}(V,T) = \int_{\infty}^{V} \left[T\left(\frac{\partial P_{\text{ex}}(V,T)}{\partial T}\right)_T - P_{\text{ex}}(V,T)\right] dV = 2RT^2 \left(\frac{\partial \eta}{\partial T}\right)_V \int_{\infty}^{V} \frac{2 + 2\eta - \eta^2}{(1-\eta)^4} \frac{1}{V} dV.$$

(16b)

In Eq. (16b), the temperature dependence of the packing fraction $(\partial \eta/\partial T)_V$ can be alternatively expressed as the linear coefficient of expansion:

$$l := \frac{1}{\sigma_{\text{EHSD}}} \left(\frac{\partial \sigma_{\text{EHSD}}}{\partial T}\right)_V = \frac{1}{3\eta} \left(\frac{\partial \eta}{\partial T}\right)_V.$$

(17)

Using $l$, the expression for the residual internal energy $U_{\text{ex}}(V,T)$ can be simplified to

$$\frac{U_{\text{ex}}(V,T)}{RT^2 l} = 6\frac{\eta^2 - 2\eta}{(1-\eta)^3}.$$

(18)

Combining Eqs. (16a) and (18), we arrive at the final expression for the residual entropy term

$$\frac{S_{\text{ex}}(V,T)}{R} = \frac{3\eta^2 - 4\eta}{(1-\eta)^2} + 6Tl\frac{\eta^2 - 2\eta}{(1-\eta)^3}.$$

(19)

Therefore, the overall thermodynamic entropy for hard spheres is expressed as the sum of the ideal gas entropy from the Sackur-Tetrode equation $S_{\text{id}}$ [Eq. (6)] and $S_{\text{ex}}$ [Eq. (19)]

$$S_{\text{hs}} = S_{\text{id}} + S_{\text{ex}}.$$

(20)

**2.4.2. Free Volume Treatment for Condensed Phases.** While Eq. (20) accounts for the hard sphere correction ($S_{\text{ex}}$) to the ideal gas entropy obtained from the Sackur-Tetrode equation ($S_{\text{id}}$), the intrinsic limitation of the Sackur-Tetrode equation implies no physical volume occupied by molecules. Therefore, even though this decomposition might provide a reasonable description of gas phase systems, it has been suggested that in condensed phases, a realistic correction to this ideal scenario is necessary for a better estimation of translational entropy.[101] In particular, building on the findings from Refs. 70 and 101, 102, we introduce the free volume approach derived from the estimated hard sphere volume in order to correct the ideal gas volume $V$ to its "effective volume," $V_{hs}$, by taking into account the influence of hard sphere repulsions. At low densities, it is reasonable to approximate $V_{hs}$ as the free volume in the system after subtracting the excluded volume:

$$V_{hs}(\rho \ll 1) = V - \left(\frac{N}{2}\right)\left(\frac{4\pi}{3}\right)\sigma^3 = V - \frac{2\pi N}{3}\sigma^3. \tag{21}$$

However, this approximation may not be valid for liquids with higher number densities. In such cases, we adopt a one-dimensional approximation for high densities based on the approach of Eyring and Hirschfelder,[102] i.e.,

$$V_{hs}(\rho \gg 0) = 8N\left[\left(\frac{V}{N}\right)^{\frac{1}{3}} - \sigma\right]^3. \tag{22}$$

Equation (22) can be intuitively understood as a three-dimensional generalization of the effective volume element calculated from the one-dimensional effective distance between the three hard sphere molecules. When both ends (1 and 3) are fixed at an averaged distance of $(V/N)^{1/3}$, the central molecule (2) is only accessible to non-repulsive regions, giving a distance of $d_{hs} = 2[(V/N)^{1/3} - \sigma]$. The effective volume $V_{hs}(\rho \gg 0)$ is then obtained as $d_{hs}^3$, resulting in Eq. (22). Finally, with this volume correction, the hard sphere entropy is expressed as a sum of the corrected ideal gas entropy using $V_{hs}$ and the hard sphere correction:

$$S_{hs} = \left[-R\ln\left(\frac{h^2}{2\pi M k_B T}\right)^{\frac{3}{2}} - R\ln\left(\frac{N}{V_{hs}}\right) + \frac{5}{2}R\right] + \left[\frac{3\eta^2 - 4\eta}{(1-\eta)^2}R + 6RTl\frac{\eta^2 - 2\eta}{(1-\eta)^3}\right]. \tag{23}$$

### 2.5. Estimating the CG Entropy: Generalized van der Waals Theory

In order to account for the attractive nature of CG interactions at large distances, the generalized van der Waals theory is formulated by defining an effective CG interaction potential using the pairwise approximation. This pairwise CG interaction, denoted as $U(R)$, gives rise to the configurational energy $U_{conf}(\mathbf{R}^N) := \sum_{I>J} U(R_{IJ})$. Alternatively, $U_{conf}$ can be expressed as a structural average over the CG ensemble

$$U_{conf}(N, V, T) = \frac{N^2}{2V}\int U(R)g(R)\,dR, \tag{24}$$

where $R = R_{IJ}$. For liquid systems, the pairwise approximation [$U(R)$ and $g(R)$] typically provides an accurate estimation of thermodynamic quantities. Yet, as discussed in Sec. 2.3, limitations in pairwise basis sets necessitate the parametrization of distinct CG potentials [$U(R)$] at different temperatures to correctly capture the temperature dependence of $U_{conf}(N, V, T)$ from the FG reference. From the definition of the CG partition function, the internal energy can be alternatively expressed as

$$U(N,V,T) = k_B T^2 \left(\frac{\partial \ln Q_{CG}}{\partial T}\right)_{N,V}.$$

(25)

Since the thermodynamic quantity in Eq. (25) is the configurational energy, $U_{\text{conf}}(N,V,T) = U(N,V,T) - U_{\text{id}}(N,V,T)$, we arrive at

$$\ln \mathbb{Z}(N,V,T) - \ln \mathbb{Z}(N,V,T=\infty) = \int_{T=\infty}^{T} \frac{U_{\text{conf}}(N,V,T)}{k_B T^2} dT$$

$$= \frac{N^2}{2V} \int_{\infty}^{T} \frac{1}{k_B T^2} \left[\int U(R) g(R) dR\right] dT.$$

(26)

For clarity, from now on we denote $\mathbb{Z}(N,V,T)$ as $\mathbb{Z}(T)$ (since there is no change in $N,V$) and $\mathbb{Z}(N,V,T=\infty)$ as the hard sphere $\mathbb{Z}_{\text{HS}}(\eta)$ (since only the hard-core repulsion term remains at $T \to \infty$). We also define the mean potential $\Phi$ as

$$\Phi := -\frac{2k_B T}{N} \int_{\infty}^{T} \frac{U_{\text{conf}}(T)}{k_B T^2} dT,$$

(27)

which results in

$$\mathbb{Z}(T) = \mathbb{Z}_{\text{HS}}(\eta) \exp\left(-\frac{N\Phi}{2k_B T}\right).$$

(28)

The thermodynamic quantities deduced from Eq. (28) can be complicated since the mean potential $\Phi$ encodes complex many-body interactions. To simplify this term further, we coarsen the attractive perturbations to a more reduced level. For example, it is reasonable to coarsen the CG interaction potential $U(R)$ to that of hard spheres with a single square-well form, $U_{\text{SW}}(R)$, as a perturbation:

$$U_{\text{SW}}(R) = \begin{cases} \infty & (\text{if } R < \sigma_{\text{EHSD}}) \\ -\epsilon & (\text{if } \sigma_{\text{EHSD}} < R < R_\epsilon \sigma_{\text{EHSD}}), \\ 0 & (\text{otherwise}) \end{cases}$$

(29)

where $R_\epsilon$ defines the width of the square-well potential. This single square-well perturbative model has been utilized in other studies,[56, 57, 103, 104] and its feasibility will be discussed in Sec. 3.3.

Using $U_{\text{SW}}(R)$, $U_{\text{conf}}(T)$ can be directly computed by defining the *attractive* coordination number $N_c := N \int g(R) \, d\mathbf{R}/V$ that considers only $\sigma_{\text{EHSD}} < R < R_\epsilon \sigma_{\text{EHSD}}$:

$$U_{\text{conf}}(T) = -\frac{N^2}{2V} \int U_{\text{SW}}(R) g(R) \, d\mathbf{R} = -\frac{N^2 \epsilon}{2V} \int_{\sigma_{\text{EHSD}}}^{R_\epsilon \sigma_{\text{EHSD}}} g(R) \, d\mathbf{R} = -\frac{N\epsilon}{2} N_c(\rho_b, T).$$

(30)

Note that $N_c$ is not a conventional coordination number but rather the coordination number originating from the attractive interactions *only,* as we assume that hard-sphere repulsions do not significantly contribute to any coordination value. Then, from Eq. (27), the mean potential becomes $\Phi = k_B T \epsilon \int_{\infty}^{T} N_c / k_B T^2 \, dT$. In principle, since $N_c$ is still dependent on temperature and directly related to $dg(T)/dT$, it is difficult to simplify $\Phi$ any further. Following its original derivation and demonstration by Sandler et al.,[72, 73] we strictly adhere to the introduction of the *low-density limit* and derive the necessary expression here, i.e., $\lim_{\rho_b \to 0} g(R; \rho_b, T) \approx \exp(-\beta U(R))$. This approximation allows us to determine the coordination number as

$$N_c = \frac{N}{V} \frac{4\pi}{3} \sigma^3 (R_\epsilon^3 - 1) \exp[\beta \epsilon], \tag{31}$$

where a detailed derivation is provided in the Appendix. Equation (31) can be an alternative starting point for reformulating Eqs. (27)-(30) in an analytical manner. If we make the further approximation that $\partial N_c(\rho_b, T)/\partial T \approx 0$, we arrive at the thermodynamic quantities expressed as

$$U_{\text{conf}} = -\frac{N \epsilon N_c}{2}, \tag{32a}$$

$$\Phi = -N_c \epsilon = \frac{2 U_{\text{conf}}}{N}. \tag{32b}$$

Therefore, under this limit, we can interpret $U_{\text{conf}}$ as the effective "averaged" interactions between neighboring molecules within the coordination shell, accounting for double-counting, while $\Phi$ represents the mean potential derived from $U_{\text{conf}}$. Altogether, the approximated configuration integral can be simplified as

$$\mathbb{Z}(T) = \mathbb{Z}_{\text{HS}}(\eta) \exp\left[-\frac{\Phi}{2 k_B T}\right]^N = \mathbb{Z}_{\text{HS}}(\eta) \exp\left[-\frac{U_{\text{conf}}}{k_B T}\right]. \tag{33}$$

We now discuss how to obtain the thermodynamic properties under the generalized van der Waals framework. Since the hard-core part of the system is faithfully mapped to $\mathbb{Z}_{\text{HS}}(\eta)$, the separability of partition functions [Eq. (5)] makes this tractable. The complete CG partition function is recovered by incorporating the momentum as

$$Q_{\text{CG}} = \frac{1}{N! h^{3N}} \mathbb{Z}_{\text{HS}}(\eta) \exp\left[-\frac{U_{\text{conf}}}{k_B T}\right] \cdot \int d\mathbf{P}^N \exp[-\beta \mathrm{T}(\mathbf{P}^N)]. \tag{34}$$

We note that the result from the previous subsection [Sec. 2.4] is valid for

$$Q_{\text{HS}}(\eta) = \frac{1}{N! h^{3N}} \mathbb{Z}_{\text{HS}}(\eta) \int d\mathbf{P}^N \exp[-\beta \mathrm{T}(\mathbf{P}^N)]. \tag{35}$$

Then, combining Eqs. (34) and (35) allows us to rewrite the CG partition function as $Q_{\text{CG}} = Q_{\text{HS}}(\eta) \exp[-U_{\text{conf}}/k_B T]$. Since the free energy is the logarithm of the partition function, this decomposition gives

$$W_{\text{CG}} = W_{\text{HS}} + U_{\text{conf}}. \tag{36}$$

Finally, the thermodynamic entropy can be formulated from $Q_{\text{CG}}$ as

$$S_{\text{CG}} = k_B \ln Q + \frac{k_B T}{Q} \frac{\partial Q}{\partial T}. \tag{37}$$

Inserting $Q_{\text{CG}}$ into Eq. (37) gives

$$S_{\text{CG}} = S_{\text{HS}} - \frac{\partial U_{\text{conf}}}{\partial T}.$$

(38)

In other words, the hard sphere entropy description can be improved by including this additional term $-\partial U_{\text{conf}}/\partial T$, which is the temperature derivative of the configurational energy originating from the attractive interactions.

## 2.6. Multiparticle expansion of the CG Entropy: Pairwise Approximation

Up to this point, we have formulated the hard sphere description of the CG entropy and progressively improved the hard sphere treatment by introducing a square-well attraction through the generalized van der Waals framework. However, these approaches are built upon several strong assumptions about the molecular CG interactions. Additionally, they require approximating several variables beforehand, e.g., $\eta$, $l$, and $U_{\text{conf}}$ for each system in order to compute the correction for the hard sphere entropy.

Alternatively, as demonstrated in pioneering work by Green, the exact formulation of the residual entropy can be derived based on the multiparticle correlation expansion without any approximation.[78, 79, 83, 105, 106] For a more in-depth discussion of residual entropy in CG modeling, readers can refer to Ref. 66. In brief, the residual entropy $S_{\text{ex}}$ is expressed as

$$S_{\text{ex}} = S - S_{\text{id}} = -\frac{1}{2}\rho \iint [g_2(\mathbf{R}_1, \mathbf{R}_2) \ln[g_2(\mathbf{R}_1, \mathbf{R}_2)] - g_2(\mathbf{R}_1, \mathbf{R}_2) + 1] d\mathbf{R}_1 d\mathbf{R}_2$$
$$-\frac{1}{6}\rho^2 \iiint g_3(\mathbf{R}_1, \mathbf{R}_2, \mathbf{R}_3) \ln[\delta g_3(\mathbf{R}_1, \mathbf{R}_2, \mathbf{R}_3)] d\mathbf{R}_1 d\mathbf{R}_2 d\mathbf{R}_3$$
$$+\frac{1}{6}\rho^2 \iiint [g_3(\mathbf{R}_1, \mathbf{R}_2, \mathbf{R}_3) - 3g_2(\mathbf{R}_1, \mathbf{R}_2)g_2(\mathbf{R}_2, \mathbf{R}_3) + 3g_2(\mathbf{R}_1, \mathbf{R}_2)$$
$$- 1] d\mathbf{R}_1 d\mathbf{R}_2 d\mathbf{R}_3 + \mathcal{O}(g_4),$$

(39)

where $g_k(\mathbf{R}_1, \cdots, \mathbf{R}_k)$ represents the $k$-body distribution functions defined by the center-of-mass coordinates of $k$ different molecules (vector). Note that we still denote the residual entropy as $S_{\text{ex}}$, where the subscript is consistent with the one used in literature as the *excess entropy*. The radial distribution function (RDF), $g(R)$, corresponds to the scalar component of the pairwise distribution: $g(R) = g_2(|\mathbf{R}_1 - \mathbf{R}_2|)$. The three-body correlation function $\delta g_3(\mathbf{R}_1, \mathbf{R}_2, \mathbf{R}_3)$ is defined as $g_3(\mathbf{R}_1, \mathbf{R}_2, \mathbf{R}_3)/g_2(\mathbf{R}_1, \mathbf{R}_2)g_2(\mathbf{R}_2, \mathbf{R}_3)g_2(\mathbf{R}_1, \mathbf{R}_3)$. By grouping contributions from $n$-particle terms, Eq. (39) can be presented in a more concise form as

$$S_{\text{ex}} = \sum_{n \geq 2} S_{\text{ex}}^{(n)},$$

(40)

where $S_{\text{ex}}^{(n)}$ involves $g_k(\mathbf{R}_1, \cdots, \mathbf{R}_k)$ terms with $k \leq n$. Typically, higher-order terms $\mathcal{O}(g_4)$ are considered negligible due to the dominance of $S_{\text{ex}}^{(2)}$ and $S_{\text{ex}}^{(3)}$. In particular, in simple liquids, it is known that $S_{\text{ex}}^{(2)}$ accounts for the majority of the residual entropy (nearly 80-90%),[83, 106-109] since the pair correlations are dominant in these cases. Therefore, at the simplest level of Eq. (39), we can approximate $S_{\text{ex}}$ by focusing solely on the translational component from $g_2(\mathbf{R}_1, \mathbf{R}_2)$,

$$S_{\text{ex}} \approx S_{\text{ex}}^{(2)} = -2\pi\rho \int_0^\infty \{g(\mathbf{R}) \ln g(\mathbf{R}) - [g(\mathbf{R}) - 1]\}\mathbf{R}^2 \cdot d\mathbf{R}.$$

(41)

We will later assess whether this pairwise description [Eq. (41)] is adequate for representing the entire residual entropy and then evaluate its performance against the (approximate) generalized van der Waals approach. Since the residual entropy only accounts for configurational phase spaces, the overall entropy approximated at the pairwise level is given as

$$S = S_{\text{id}} + S_{\text{ex}}^{(2)}. \tag{42}$$

## 2.7. Computational Details

Based on previous studies of CG entropy in liquid systems, we select the same molecular systems for this study: methanol and chloroform.[43] In Ref. 43, both liquids exhibit a liquid phase at a temperature of 300 K, and we chose this same temperature as our target temperature for this work. Our computational protocol closely follows the protocol used in our prior studies. To outline the procedure, we initially generated the FG configurations. These configurations consisted of 1000 molecules and were created using the Packmol program package[110], and the topology was generated by the Visual Molecular Dynamics (VMD) program.[111] Starting from these initial configurations, we proceeded to relax structures by employing the energy minimization techniques, and we gradually annealed the system up to 300 K with the Nosé-Hoover thermostat[112, 113] using $\tau_{\text{NVT}} = 0.1$ ps over 0.1 ns. We then equilibrated the target system under constant *NPT* dynamics at 1 atm using $\tau_{\text{NPT}} = 1.0$ ps, where the Andersen barostat[114] was used. Finally, we sampled the constant *NVT* configurations at every 1 ps over a duration of 5 ns of additional production dynamics steps using the Nosé-Hoover thermostat for constructing the CG model. Nevertheless, several different temperatures are needed to be sampled to calculate temperature dependent properties, such as $l$ defined in Eq. (17). The additional computational details for obtaining the $l$ values will be discussed in Sec. 3.2.2-3.2.3.

From the produced constant *NVT* runs at the FG level, we constructed the CG model by employing a center-of-mass mapping. From the manually mapped FG trajectories, the effective CG interactions were parametrized using 6$^{\text{th}}$ order B-splines at a resolution of 0.20 Å. We used the publicly available open-source software OpenMSCG.[115] In order to enhance the sampling of the inner-core region, we additionally fitted the polynomial to a form $A \cdot r^{-B}$ at short distances.[116] Using the parametrized CG interactions, the CG simulations were performed under constant *NVT* dynamics using the Nosé-Hoover thermostat for 5 ns, and we collected the CG configuration every 1 ps.

## 3. Results
## 3.1. Ideal Gas Approximation

The estimation of the CG entropy in the ideal gas approximation can be readily made utilizing the Sackur-Tetrode equation [Eq. (6)]. We applied this equation to methanol and chloroform. Table 1 compares the estimated CG entropy with the actual CG entropy computed using the two-phase thermodynamic (2PT) method for the CG trajectories. In brief, the 2PT methodology can efficiently compute the thermodynamic properties of a given system in a relatively short time (~ 20 ps) while maintaining near-identical accuracy compared to other conventional free energy methodologies.[117, 118] Reference 118 provides a detailed comparison of the 2PT method with other free energy approaches to estimate thermodynamic properties.

Remarkably, the ideal gas entropy provides values of a similar order of magnitude as the reference CG entropy. While the entropy values are overestimated by about $1.5 - 4$ cal·mol$^{-1}$·K$^{-1}$ for both systems, we believe that this agreement suggests that the ideal gas approximation can be a reasonably useful method for a rough estimation of CG entropy, suitable for quick back-of-the-envelope calculations. In other words, since the Sackur-Tetrode equation only requires information about the system density and molecular characteristics, the ideal gas description can serve as an initial check to provide a rough estimate of the CG entropy, offering insight into the mapping entropy. The overestimated value of the entropy in the ideal gas description can be attributed to its definition, where an ideal gas is assumed to occupy all available configurations, resulting in higher entropy values compared to molecular systems with non-zero interactions.

**Table 1.** Thermodynamic entropy for methanol and chloroform estimated using the ideal gas approximation in comparison with the reference entropy values obtained by 2PT calculations from CG trajectories.

| Molecule | Thermodynamic Entropy Quantity (cal·mol$^{-1}$·K$^{-1}$) | |
|---|---|---|
| | Actual CG Entropy | Ideal gas estimation |
| Methanol | 21.110 | 23.710 |
| Chloroform | 24.938 | 28.894 |

## 3.2. Hard Sphere Theory

*3.2.1. Hard Sphere Mapping.* To enhance the ideal gas description, we now map the CG system to an effective hard sphere system. The estimation of $\sigma_{BH}$ was performed using Eq. (9) with the parametrized CG PMFs, which were obtained through force matching and are illustrated in Fig. 1.

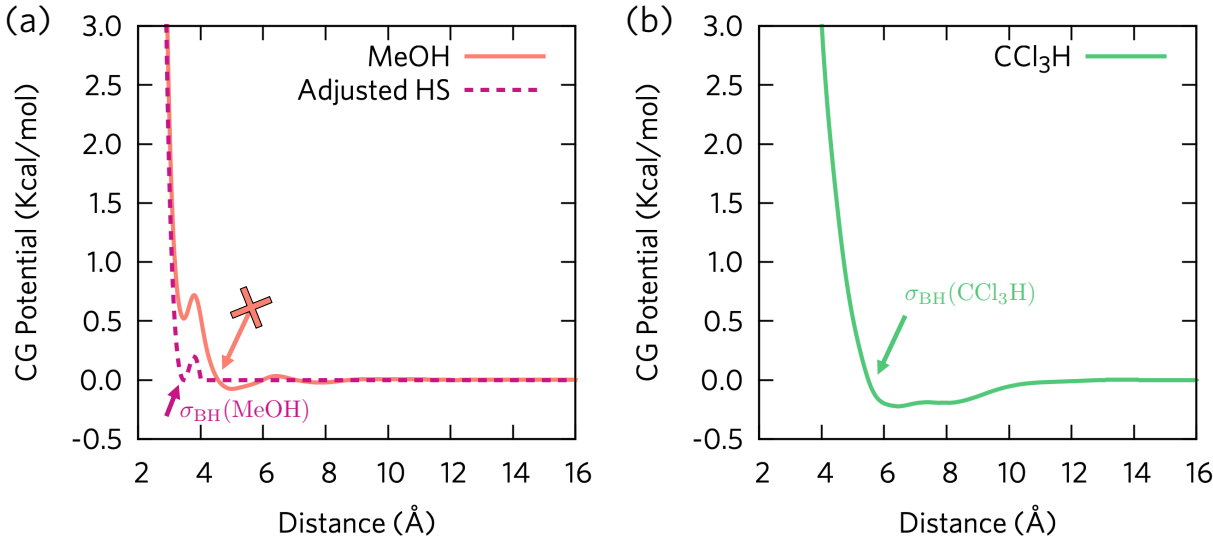

**Figure 1.** Pairwise CG interactions obtained from force matching for (a) methanol and (b) chloroform at 300 K. The corresponding hard sphere diameters are marked with arrows. Note that the additional repulsive ordering in methanol results in an unphysical Barker-Henderson diameter (red arrow), and this overestimated diameter can be corrected by employing the Weeks-Chandler-Andersen perturbation theory [Eq. (42)], which gives correct repulsive behavior at short distances (magenta arrow).

Note that the methanol interactions depicted in Fig. 1(a) are slightly different from conventional hard sphere systems. Unlike traditional hard sphere systems, the effective CG interactions here do not monotonically decay to zero after the short-range hard-core regime. This behavior in methanol can be attributed to additional structuring caused by its non-trivial interactions, as observed in its RDF. In this context, directly applying Eq. (9) yields a nonphysical hard sphere diameter value, as it considers contributions beyond $R > R_0$, resulting in a value of 4.54 Å. To address this issue, we use the Weeks-Chandler-Andersen approach[58-60] to adjust this repulsive region by shifting the CG interactions according to

$$U^{\text{HS}}(R) = U(R) - U(R_{\min}), \quad (43)$$

where $R_{\min}$ represents the distance where the effective force is near zero, i.e., $R_{\min}$ = 3.46 Å, as observed in Fig. 1(a). This adjustment is often referred to as the hybrid Barker-Henderson approximation in the literature.[119, 120] Using this shifted CG interaction, depicted as dashed lines in Fig. 1, we obtained an adjusted Barker-Henderson diameter of 3.119 Å. This value is much closer to $R_{\min}$ and is expected to more accurately represent the hard sphere nature of methanol. Nevertheless, the discrepancy between the Barker-Henderson and Weeks-Chandler-Andersen approaches implies that employing the hard sphere description for estimating methanol's thermodynamic properties may not be straightforward. We will investigate this further in Sec. 3.

Conversely, for chloroform, no such pathologies are observed in the estimate of CG interactions. This is consistent with the relatively spherical profile of chloroform when compared to methanol, giving a Barker-Henderson diameter of 4.988 Å. From the obtained hard sphere diameters, we can relate them to the effective packing fraction $\eta_{\text{BH}}$ through the definition $\eta_{\text{BH}} = \frac{\pi}{6}\sigma_{\text{BH}}^3 \rho_b$.

***3.2.2. Temperature Dependence of Hard Sphere Diameter: Proof-of-concept.*** In order to correctly account for the hard sphere contribution to the CG entropy, one must determine the temperature dependence of the hard sphere diameter, denoted as $l \coloneqq (\partial \sigma_{\text{BH}}/\partial T)_V/\sigma_{\text{BH}}$. Since $l$ is defined under constant volume conditions, we constructed the CG systems at different temperatures while maintaining the equilibrium volume at 300 K, our target temperature.

Since $\sigma_{\text{BH}}$ encodes the many-body nature exhibited by the many-body CG PMF $U(R)$, the analytical determination of $\sigma_{\text{BH}}$ and $l$ is practically impossible. Instead, we compute $\sigma_{\text{BH}}$ values over a wide temperature range and approximate the partial derivative through finite differences. It is noteworthy that, to our knowledge, such an attempt to compute the temperature response of the hard sphere diameter extracted from CG systems has not been reported in the literature. Thus, we first validate our approach with a proof-of-concept system. A similar proof-of-concept study was proposed to investigate the nature of the pairwise entropy contribution from the CG PMF to the CG entropy in Ref. 43.

To ensure the isotropic symmetry of the CG site resulting from the center-of-mass mapping, we constructed four tetrahedral-like structures based on CCl$_4$ while varying the non-bonded interactions between the FG particles. We first generated a system, which we call X$\alpha_4$, where the non-bonded interactions are described using the Lennard-Jones (LJ) interaction with harmonic bonded interactions. Then, we modified one of the X-α pair interactions to be much weaker and shorter, i.e., from σ = 4.5 Å to 3.0 Å and ε = 0.32 kcal/mol to 0.2 kcal/mol, as detailed in Table 2. For these four altered systems (X$\alpha_4$, X$\alpha_3\beta$, X$\alpha_2\beta\chi$, and X$\alpha\beta\chi\delta$), the bonded interactions between X and other atoms are modeled as $d$ = 1.90 Å and $k$ = 200 kcal/mol. In order to solely consider LJ interactions with harmonic bonds, we set the charges on each atom to zero.

**Table 2**. Imposed non-bonding interaction parameters (LJ σ and ε) for the proof-of-concept systems based on the tetrahedral motif (Xα4), where we altered X-α interactions in order to introduce heterogeneously bonded molecular systems.

| System | LJ coefficient | Non-bonding interaction | | | | |
|---|---|---|---|---|---|---|
| | | X-α | X-δ | X-χ | X-β | X-X |
| Xα4 | σ (Å) | | 4.5 | | | 2.50 |
| | ε (kcal/mol) | | 0.32 | | | 0.05 |
| Xα3β | σ (Å) | | 4.5 | | 3.0 | 2.50 |
| | ε (kcal/mol) | | 0.32 | | 0.2 | 0.05 |
| Xα2βχ | σ (Å) | | 4.5 | 3.5 | 3.0 | 2.50 |
| | ε (kcal/mol) | | 0.32 | 0.24 | 0.2 | 0.05 |
| Xαβχδ | σ (Å) | 4.5 | 4.0 | 3.5 | 3.0 | 2.50 |
| | ε (kcal/mol) | 0.32 | 0.28 | 0.24 | 0.2 | 0.05 |

We conducted the FG simulations of these systems and constructed the CG models using the MS-CG methodology. These CG trajectories were propagated at temperatures of 225, 250, 275, and 300 K, where all systems exhibit liquid-like behavior. From the parametrized CG interactions, the Barker-Henderson diameters for the test systems were readily determined, as shown in Fig. 2.

Upon introducing shorter non-bonded pair interactions, we observed a gradual decrease in the $\sigma_{BH}$ values from Xα4 to Xαβχδ. Also, for all four systems, $\sigma_{BH}$ decreased at higher temperatures, in line with the typical behavior of hard spheres, where each hard sphere can overcome repulsions more readily at higher temperatures. The monotonically decreasing trend of $\sigma_{BH}$ with respect to temperature in Fig. 2(a) allows us to compute the temperature derivative term. Figure 2(b) compares different $l$ values across the test systems at 300 K, evaluated via finite difference, revealing that the monotonic trend persists, with the magnitude of $l$ increasing as the system becomes less strongly bound. This aligns with our design principle: the $\sigma_{BH}$ value of Xα4 should be less flexible than that of Xαβχδ because the altered interactions are relatively stronger than other X-β (χ or δ). Consequently, we find that the hard sphere diameter obtained from the Barker-Henderson approach provides a reasonable description of the test systems, with their temperature responses reasonably captured using the finite difference method.

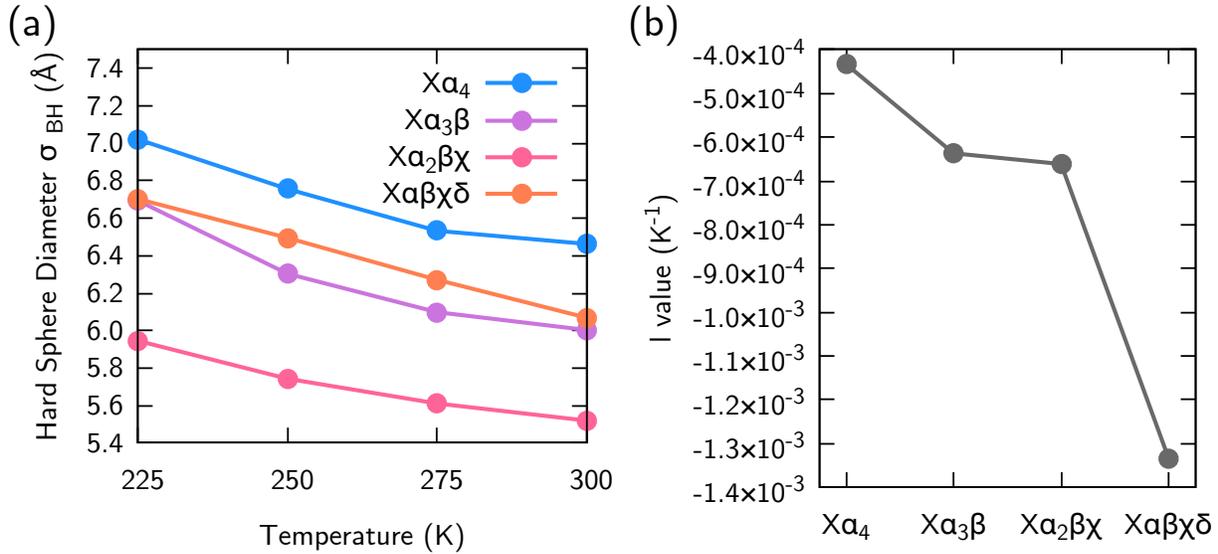

**Figure 2**. Temperature dependence of the hard sphere diameters from the prototypical tetrahedral systems: Xα$_4$ (blue), Xα$_3$β (purple), Xα$_2$βχ (red), and Xαβχδ (orange). (a) Hard sphere diameters $\sigma_{BH}$ for four test systems at different temperatures. (b) Temperature derivative of $\sigma_{BH}$ at 300 K for the four test systems.

*3.2.3. Temperature dependence of Hard sphere diameter: Molecular Liquids.* In order to calculate $l$ for molecular liquids, we considered the following temperature ranges: 300-400 K at intervals of 25 K for methanol and 250-350 K at intervals of 25 K for chloroform. Note that similar to the test systems, our target temperature was set at 300 K, and the volume was held constant, resulting in a cubic box length of 41.301 Å for methanol and 51.050 Å for chloroform.

The $\sigma_{BH}$ values for the molecular liquids were subsequently determined by employing Eq. (9) for each PMF. For methanol, the Weeks-Chandler-Andersen treatment based on Eq. (43) was applied, as depicted in Fig. 3. Like the test systems, we observed a monotonically decreasing trend in $\sigma_{BH}$ with respect to temperature for both liquids. Notably, this trend appears almost linear for all temperature conditions, which is expected from previous studies on hard spheres. This linearity also indicates the numerical stability of estimating $l$ using finite differences. It is worth noting that the non-linear trend seen in Fig. 2 might be due to the test systems not precisely representing the real molecular liquids, particularly in terms of missing charge interactions. Nevertheless, from the linearly-varying feature of $\sigma_{BH}$ in methanol and chloroform, we computed the temperature derivative of $\sigma_{BH}$ value, yielding

$$l_{\text{MeOH}} = \frac{1}{\sigma_{BH}}\left(\frac{\partial \sigma_{BH}}{\partial T}\right)_V = \frac{1}{3.119\text{Å}} \cdot (-4.944 \times 10^{-3}\text{Å} \cdot \text{K}^{-1}) = -1.585 \times 10^{-3}\text{K}^{-1},$$

(44a)

$$l_{\text{CCl}_3\text{H}} = \frac{1}{\sigma_{BH}}\left(\frac{\partial \sigma_{BH}}{\partial T}\right)_V = \frac{1}{4.988\text{Å}} \cdot (-6.964 \times 10^{-3}\text{Å} \cdot \text{K}^{-1}) = -1.396 \times 10^{-3}\text{K}^{-1}.$$

(44b)

The obtained $l$ values fall within the range derived by Ben-Amotz and Herschbach.[121]

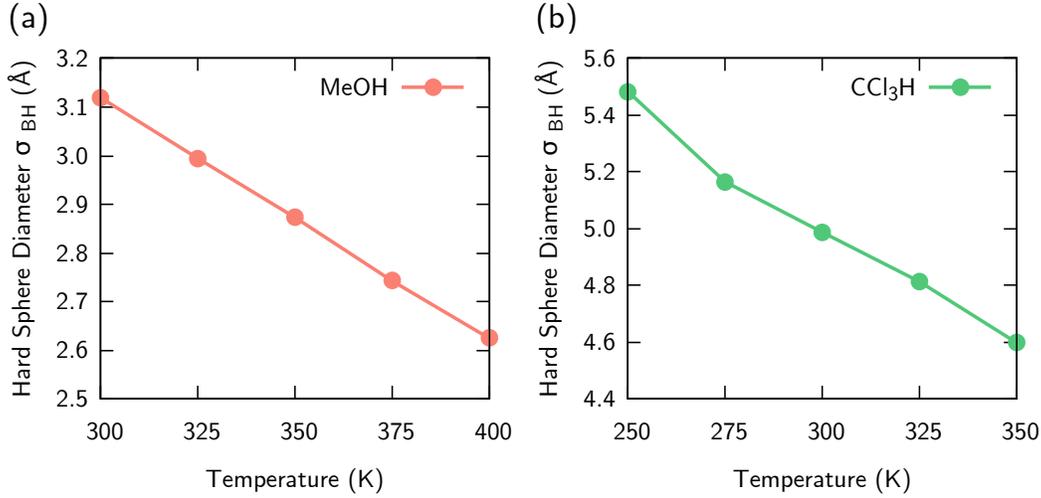

**Figure 3.** Temperature dependence of the hard sphere diameters at different temperatures: (a) methanol (red) and (b) chloroform (green).

Note that the packing fraction for each system can also be determined from $\sigma_{BH}$ and the number density,

$$\eta_{\text{MeOH}} = \frac{\pi}{6} \cdot (3.119 \text{ Å})^3 \cdot \frac{1000}{(41.301 \text{ Å})^3} = 0.2256, \tag{45a}$$

$$\eta_{\text{CCl}_3\text{H}} = \frac{\pi}{6} \cdot (4.988 \text{ Å})^3 \cdot \frac{1000}{(51.050 \text{ Å})^3} = 0.4883. \tag{45b}$$

We can now calculate each term in Eq. (23) to evaluate the CG entropy using the hard sphere description. As discussed earlier, in the presence of strong repulsion from the hard-core at short distances, an ideal gas contribution needs to be corrected using Eq. (23). For methanol, both the free volume and one-dimensional approximations yield similar corrected volume terms, with $V_{\text{hs}}(\rho \ll 1) = (19.023 \text{ Å})^3$ and $V_{\text{hs}}(\rho \gg 0) = (20.216 \text{ Å})^3$. However, for chloroform, we observe that the naïvely estimated free volume approximation does not hold: $V_{\text{hs}}(\rho \ll 1)$ gives a value of $-1.174 \times 10^{-25} \text{ Å}^3$, indicating an overestimation of the excluded volume. This discrepancy can be attributed to the high packing fraction of chloroform, which is more than twice that of methanol. While the free volume approximation fails in this case, Eq. (22) can correctly describe the high-density behavior of chloroform, yielding $V_{\text{hs}}(\rho \gg 0) = (4.7125 \text{ Å})^3$.

*3.2.4. Results: Thermodynamic Entropy.* The corrected ideal gas entropy values are presented in Table 3. As expected, the changes upon including hard sphere repulsions are much more pronounced in chloroform, with a difference of 10.248 cal·mol$^{-1}$·K$^{-1}$, whereas the changes in the methanol value are relatively smaller, approximately 1.659 cal·mol$^{-1}$·K$^{-1}$. While the ideal gas entropy term itself decreased significantly, the hard sphere correction term compensates for this entropy loss. For methanol, this correction term was determined to be 2.418 cal·mol$^{-1}$·K$^{-1}$, resulting in a final entropy value of 21.506 cal·mol$^{-1}$·K$^{-1}$. When compared to the actual CG entropy, the absolute error is within half a cal·mol$^{-1}$·K$^{-1}$, indicating that the hard sphere description significantly enhances entropy estimation.

**Table 3.** Thermodynamic entropy for methanol and chloroform estimated using the hard sphere approach in comparison with the reference entropy values obtained by 2PT calculations from CG trajectories. Note that ideal gas contributions are slightly different from those in Table 1 due to changes in volume, suggested by Eq. (22).

| Molecule | Thermodynamic Entropy Quantity (cal·mol$^{-1}$·K$^{-1}$) | | | |
| --- | --- | --- | --- | --- |
| | Actual CG Entropy | Hard Sphere Approach | | |
| | | Volume-corrected ideal gas term | Hard sphere correction | Overall |
| Methanol | 21.110 | 19.451 | 2.418 | 21.506 |
| Chloroform | 24.938 | 14.690 | 13.839 | 28.530 |

We arrived at a similar conclusion for chloroform. While more than 10 cal·mol$^{-1}$·K$^{-1}$ are lost in the ideal gas term due to hard sphere repulsions, the hard sphere correction term is larger than this loss, yielding a final value of 28.530 cal·mol$^{-1}$·K$^{-1}$. However, as expected given the high-density condition of chloroform, the hard sphere correction only marginally enhances the ideal gas description. The relative error for chloroform decreases from 15.86 % to 14.40 %, while for methanol, the relative error decreases significantly from 12.32 % to 1.877 %.

In summary, we find that the CG entropy can be reasonably described by the hard sphere model derived from classical perturbation theory. While this approach necessitates the inclusion of several additional parameters beyond solely the hard sphere diameter, it allows us to avoid introducing any stronger assumptions and effectively recapitulates the entropy at relatively low densities, outperforming the ideal gas description. Overall, we believe that this approach holds promise for efficiently estimating the entropy of complex molecular CG liquids.

### 3.3. Generalized van der Waals Theory
*3.3.1. Mapping the Attractive Interaction.* In Table 3, we observe that the hard sphere entropies still exceed the reference CG entropies, and one might expect that considering the attractive nature of CG interactions could help mitigate this overestimation. As described in Sec. 2.5, we now explore this additional effect by including the perturbative attractions within a generalized van der Waals framework. This framework allows us to approximate the attractive part of the effective CG interactions as a step potential.

Repulsive cores coupled with attractive stepwise potentials have been widely utilized to model liquids and their phase transitions.[56, 57, 103, 104] In particular, the width and depth of a step potential plays a critical role in influencing phase transitions, as seen in pressure-temperature phase diagrams. A notable approach in this context is the Stell-Hemmer Hamiltonian, which originates from the occupied and unoccupied cells of a lattice gas model. This Hamiltonian introduces core-softened potentials, effectively capturing second critical points.[122, 123] Furthermore, core-softened potentials with long-range attractions have been successfully employed to investigate liquid systems, e.g., water, at the atomistic level.[124-126] Given the extension of core-softened potentials to CG water,[127-129] it is reasonable to expect that approximating long-range attractions as step potentials should yield reasonably accurate results for systems such as the molecular CG liquids pursued in this work.

We note that the generalized van der Waals approach, as outlined in Eqs. (26)-(32), was derived under the assumption of the low-density limit [Eq. (31)].[130] To appropriately address this assumption, the square-well interaction is estimated from the pair potential of mean force (PPMF),[131] which can be obtained by the FG RDF using the inverse work theorem: $U_{\text{pair}}(R) := -k_B T \ln g(R)$. It is important to highlight that the PPMF does not necessarily reflect the true physical attraction between CG molecules, as it includes environment-mediated forces along with pair interactions. A comprehensive discussion and analysis of the non-trivial contributions of higher-order environment interactions in CG interactions can be found in Refs. 132-134, and more recently in Ref. 135. However, in this section, we closely adhere to the low-density limit proposed by Refs. 72 and 130. From the obtained $U_{\text{pair}}(R)$ at different temperatures, we extracted the well depth parameter, denoted as $\epsilon$. Table 4 lists the $\epsilon$ values computed based on the pair contributions of the methanol and chloroform CG interactions.

**Table 4.** Mapped well depths and their mean temperature derivative computed by finite difference of the CG interactions of (a) methanol and (b) chloroform.

| (a) Methanol | | |
|---|---|---|
| Temperature | $\epsilon$ value (cal/mol) | $\Delta\epsilon/\Delta T$ (cal/mol/K) |
| 300 K | -190.421 | |
| 325 K | -208.609 | |
| 350 K | -228.539 | -0.817 |
| 375 K | -249.540 | |
| 400 K | -272.073 | |

| (b) Chloroform | | |
|---|---|---|
| Temperature | $\epsilon$ value (cal/mol) | $\Delta\epsilon/\Delta T$ (cal/mol/K) |
| 300 K | -315.977 | |
| 325 K | -343.963 | |
| 350 K | -371.565 | -1.124 |
| 375 K | -399.603 | |
| 400 K | -428.348 | |

***3.3.2. Results: Thermodynamic Entropy.*** Next, we estimate the effective coordination number due to the attraction, $N_c$, from the FG RDF by numerically computing

$$N_c \approx \rho \int_{R_1}^{R_2} 4\pi R^2 g(R) dR,$$

(46)

where $R_1$ and $R_2$ correspond to the leftmost and rightmost distances, respectively, in the square well potential. In practice, these distances are determined from the minimum and maximum distances in the attractive regime where the pair contribution to the CG PMF is zero: $U_{\text{pair}}(R_1) = U_{\text{pair}}(R_2) = 0$ and $R_1 < R_2$. Even though Eq. (46) accounts for the contributions from long-range attractions, the estimation of $N_c$ relies on utilizing $g(R)$ from the FG system in a bottom-up manner. Hence, we anticipate some deviations due to the complex interaction profiles embedded in the FG $g(R)$. This may have a more significant impact on non-hard sphere systems, such as methanol. These distances were computed to be $R_1 = 4.04$ Å and $R_2 = 5.36$ Å for methanol, and $R_1 = 4.58$ Å and $R_2 = 6.52$ Å for chloroform.

In order to estimate the thermodynamic entropy using the generalized van der Waals approach in a tractable form, Eq. (32) assumed $N_c$ to be independent of temperature, and hence we estimated at the target temperature of 300 K for both methanol and chloroform. With the estimated $\epsilon$ and $N_c$, we then computed approximate $U_{\text{conf}}$ values at different temperatures. The temperature derivative of $U_{\text{conf}}$, denoted as $\partial U_{\text{conf}}/\partial T$, was calculated using finite differences. With these introduced approximations in Eqs. (26)-(32), we observe that $\partial U_{\text{conf}}/\partial T$ is essentially identical to

$$\frac{\partial U_{\text{conf}}}{\partial T} = -\frac{NN_c}{2}\frac{\partial \epsilon}{\partial T} \approx -\frac{NN_c}{2}\frac{\Delta \epsilon}{\Delta T}. \tag{47}$$

Notably, the $\epsilon$ values obtained for methanol and chloroform at different temperature intervals, as reported in Table 4, indicate a nearly constant $\Delta\epsilon/\Delta T$ under these conditions. Unlike the effective CG interaction, where its temperature derivative becomes the effective entropic contribution, $\epsilon$ in Eq. (47) represents the PPMF, making the physical meaning underlying $\Delta\epsilon/\Delta T$ is less clear. Nevertheless, it is interesting to note the nearly identical $\Delta\epsilon/\Delta T$ values for the liquids studied here, and further investigation in future work along this line is expected to provide a clearer understanding of the microscopic origins underlying the PPMF and its temperature variation.

**Table 5.** Thermodynamic entropy for methanol and chloroform estimated using the generalized van der Waals approach in comparison with the reference entropy values obtained by employing 2PT calculations on actual CG trajectories. Note that for the generalized van der Waals approach, we only listed the correction factor $\partial U_{\text{conf}}/\partial T$ that can be combined with the results from Table 3 to estimate the overall estimated entropy values.

| Molecule | Thermodynamic Entropy Quantity (cal·mol$^{-1}$·K$^{-1}$) | | | |
|---|---|---|---|---|
| | Actual CG Entropy | Hard Sphere Approach | Generalized van der Waals approach | |
| | | | Correction | Overall |
| Methanol | 21.110 | 21.506 | -1.301 | 20.206 |
| Chloroform | 24.938 | 28.530 | -2.398 | 26.132 |

Finally, the correction factor from the generalized van der Waals theory was computed using

$$S_{\text{g-vdW}}^{\text{corr}} = -\frac{\partial U_{\text{conf}}}{\partial T} \approx -\frac{\Delta U_{\text{conf}}}{\Delta T} \approx \frac{NN_c}{2}\frac{\Delta \epsilon}{\Delta T}, \tag{48}$$

as presented in Table 5. In both cases, we observe that the generalized van der Waals treatment provides a negative correction to the hard sphere reference. This is expected due to the attractive nature predicted by the generalized van der Waals theory according to Eq. (48). However, due to the non-hard sphere nature of methanol, the generalized van der Waals approach slightly underestimates the attractive contribution by 0.9 cal·mol$^{-1}$·K$^{-1}$ compared to the reference CG entropy value. In contrast, the correction for chloroform is greater in value, with the final corrected entropy that is in closer agreement with the reference value compared to the hard sphere entropy reported in Table 3. The van der Waals treatment thus reduces the discrepancy from 14.40 % to 4.79 %. Based on these findings in Table 5, we conclude that the generalized van der Waals approach can rectify the overly estimated entropy values from the hard sphere nature by accurately addressing the attractive part of the CG interactions. Nevertheless, given the simplified nature of the square-well interaction, this framework may be more suitable to systems exhibiting hard sphere characteristics akin to chloroform. In this light, we note that an additional improvement to this approach is possible in which the attraction can be generalized to a more complicated shape beyond the square-well interaction form.

### 3.4. Multiparticle Expansion
#### 3.4.1. Pairwise Contribution: CG System.
Unlike the aforementioned approaches, an alternative expression rooted in the statistical mechanical definition of residual entropy can circumvent the limitations posed by model assumptions, such as hard sphere or generalized van der Waals theory. Equation (39) asserts that these limitations are not necessary to estimate the overall thermodynamic entropy of the CG system as long as one has correct structural correlations of the target system. Nevertheless, the aforementioned hard sphere and generalized van der Waals theory can be effectively folded into Eq. (39) as well, such that these two models give rise to approximate correlation functions such as $g_2(\mathbf{r}_1, \mathbf{r}_2)$ and $g_3(\mathbf{r}_1, \mathbf{r}_2, \mathbf{r}_3)$. Therefore, the multiparticle expansion is expected to further enhance the fidelity of entropy estimation by directly employing more accurate correlation functions.

Even though an accurate determination of Eq. (39) requires these correlation functions from the CG simulations, the bottom-up nature of CG models asserts that there is no need to run CG simulations, since the CG correlation functions can be approximated using the FG references. For example, if we aim to calculate the pairwise contributions only [Eq. (41)], the residual entropy portion can be readily estimated from the FG statistics, given that pairwise distribution functions are well-captured using the MS-CG methodology. Additionally, the single-site CG model greatly benefits from evaluating Eq. (41) since, at the CG resolution, the $\mathbf{R}$ vector is equivalent to the magnitude of $\mathbf{R}$, $|\mathbf{R}| = R$. This is because the CG pair distances do not contain any orientational information, which is already integrated out during the CG process. We note that a correct estimation of the FG entropy is highly challenging even for the pairwise contribution because one needs to calculate the orientational contribution, $S_{\text{or}}^{(2)}$, of molecular pairs

$$S_{\text{or}}^{(2)} = -2\pi\rho \int_0^\infty g^{(2)}(R) \cdot \left(-\frac{1}{\Omega^2} \int \int \mathcal{J}(\omega_1, \omega_2) g(\omega_1, \omega_2|R) \ln g(\omega_1, \omega_2|R) d\omega_1 d\omega_2\right) dR, \tag{49}$$

where $\omega_1, \omega_2$ are the angular variables of the molecular pair, and $\mathcal{J}(\omega_1, \omega_2)$ denotes the Jacobian associated with the angular variables. For homogeneous liquids, Eq. (49) reduces to the integration of five[136] or six angular variables.[137] Hence, while numerical integration is generally highly challenging, even for small molecules (e.g., water),[137-140] this is not the case here. Using the FG RDF, we can estimate $S_{\text{ex}}^{(2)}$ for the single-site CG system as

$$S_{\text{ex}}^{(2)} = -2\pi R\rho \int_0^\infty \{g(R) \ln g(R) - [g(R) - 1]\} R^2 \cdot dR, \tag{50}$$

without the need for additional CG simulations.

#### 3.4.2. Pairwise Contribution: Results.
Inserting Eq. (50) into Eq. (42), we arrive at the overall CG entropy

$$S^{(2)} = \left[-R\ln\left(\frac{h^2}{2\pi M k_B T}\right)^{\frac{3}{2}} - R\ln\left(\frac{N}{V}\right) + \frac{5}{2}R\right]$$
$$+ \left[-2\pi R\rho \int_0^\infty \{g(r)\ln g(r) - [g(r)-1]\}r^2 \cdot dr\right].$$
(51)

In Eq. (51), the ideal gas contribution $S_{id}$ does not need to be adjusted based on the hard sphere volume, as the entropy correction term effectively accounts for the residual entropy originating from the ideal contribution. Therefore unlike the situations with hard sphere and generalized van der Waals models, we can systematically incorporate the correction term directly into the values presented in Table 1. However, as noted earlier, Eq. (51) only considers the pairwise contributions to the residual entropy.

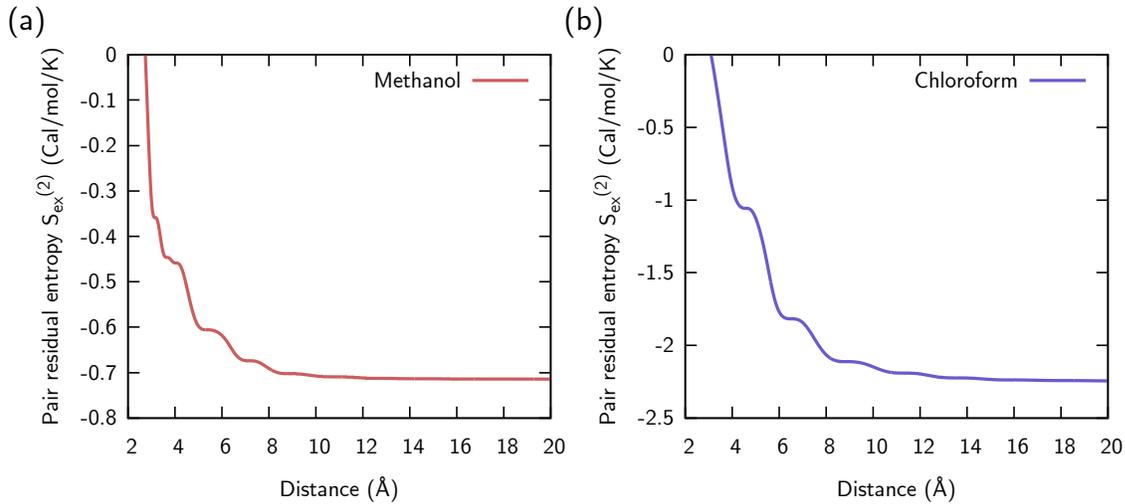

**Figure 4.** Computed pair residual entropy for (a) methanol and (b) chloroform. We depict $S_{ex}^{(2)}(R)$ as a function of the pair distance $R$, which is the upper limit of the integration of Eq. (50). The general profile is consistent with the reported behavior from Ref. 141. The converged values are listed in Table 6.

Ideally, the total residual entropy should be smaller than $S^{(2)}$ due to the negative value contributed by higher-order residual entropy (note that the residual entropy is always a negative quantity). Under the pairwise approximation, the contribution from this higher-order entropy term can be obtained through an approximation proposed by Laird and Haymet[142, 143]

$$S_{\text{pair}} = S^{(2)} - \frac{1}{2}\mathcal{R},$$
(52)

where $\mathcal{R}$ is a gas constant. Since both methanol and chloroform can be regarded as pair-dominant liquids, we also apply Eq. (52) to see if the pairwise approximation could estimate the CG entropy term correctly. As shown in Fig. 4, numerical integrations of $S_{ex}^{(2)}$ were performed over the radial domain, and we utilized the in-house code employed in Ref. 43 to estimate the correction for the entropy. $S_{ex}^{(2)}$ and $S_{\text{pair}}$ from the FG RDF at 300 K were chosen as the converged values at large distances, as shown in Fig. 4 and listed in Table 6.

**Table 6.** Thermodynamic entropy for methanol and chloroform estimated using the multiparticle expansion approach in comparison with the reference entropy values obtained by employing 2PT calculations on actual CG trajectories. Here, the systematic expansion up to the second order, $S_{\text{pair}}$, was considered with the additional correction term given by Eq. (52).

| Molecule | Thermodynamic Entropy Quantity (cal·mol$^{-1}$·K$^{-1}$) | | | |
| --- | --- | --- | --- | --- |
| | Actual CG Entropy | Ideal gas estimation | Pairwise correction | $S_{\text{pair}}$ |
| Methanol | 21.110 | 23.710 | -0.714 | 22.002 |
| Chloroform | 24.938 | 28.894 | -2.244 | 25.656 |

As previously discussed, the contribution solely from the ideal gas approximation consistently overestimates the entropy, because there are no interactions between particles. While the pairwise correction $S_{\text{ex}}^{(2)}$ can significantly enhance this description, having only $S_{\text{ex}}^{(2)}$ value still results in a slight overestimation of the overall CG entropy. For methanol, $S_{\text{id}} + S_{\text{ex}}^{(2)}$ gives 22.996 cal·mol$^{-1}$·K$^{-1}$, which exceeds the actual CG entropy by 1.886 cal·mol$^{-1}$·K$^{-1}$. A similar situation arises with chloroform, where $S_{\text{id}} + S_{\text{ex}}^{(2)}$ exceeds the reference value by 1.712 cal·mol$^{-1}$·K$^{-1}$. By approximating the remaining higher-order entropy contribution as $-\mathcal{R}/2$, roughly 0.994 cal·mol$^{-1}$·K$^{-1}$, the final pairwise CG entropy offers much more accurate values. As displayed in Table 6, $S_{\text{pair}}$ estimates the CG entropy within 1 cal·mol$^{-1}$·K$^{-1}$ in comparison to the references for both liquids. Nevertheless, $S_{\text{pair}}$ is slightly larger than the actual CG value, suggesting that explicitly calculating the higher-order entropy contributions might be needed to reduce the overestimated pairwise entropy.

***3.4.3. Three-body Contribution.*** The next step in the systematic framework is to explicitly include the three-body contribution. The three-body correction for the CG entropy can be extracted from Eq. (39), originally derived by Wallace and Evans.[79-82] Note that the three-body correction shown in Eq. (39) assumes the symmetry over the triplets, i.e., $S_{\text{ex}}^{(3)}$ can be written as

$$S_{\text{ex}}^{(3)} = -\frac{1}{6}\rho^2 \iiint g_3(\mathbf{R}_1, \mathbf{R}_2, \mathbf{R}_3) \ln[\delta g_3(\mathbf{R}_1, \mathbf{R}_2, \mathbf{R}_3)] \, d\mathbf{R}_1 d\mathbf{R}_2 d\mathbf{R}_3$$
$$+ \frac{1}{6}\rho^2 \iiint [g_3(\mathbf{R}_1, \mathbf{R}_2, \mathbf{R}_3) - 3g_2(\mathbf{R}_1, \mathbf{R}_2)g_2(\mathbf{R}_2, \mathbf{R}_3) + 3g_2(\mathbf{R}_1, \mathbf{R}_2) - 1] d\mathbf{R}_1 d\mathbf{R}_2 d\mathbf{R}_3.$$

(53)

Ideally, one can generate the three-body histograms from FG statistics and compute the three-body correlations, which would provide an accurate estimate for $S_{\text{ex}}^{(3)}$. Alternatively, several papers have also reported computing Eq. (53) by defining an $H$ function based on the work by Egelstaff[144]

$$H(R, S, T) = g_3(R, S, T) - g_2(R)g_2(S)g_2(T)$$

(54)

where $R = |\mathbf{R}| = |\mathbf{R}_1 - \mathbf{R}_2|$, $S = |\mathbf{S}| = |\mathbf{R}_2 - \mathbf{R}_3|$, and $T = |\mathbf{R} - \mathbf{S}|$. An analytical advantage of $H$ is that its Fourier-transformed counterpart $\widetilde{H}(Q)$ decouples the triple correlations, and $\widetilde{H}(Q)$ can be computed in reciprocal space $Q$.[105] Yet, obtaining a converged $\widetilde{H}(Q)$ requires sufficient sampling and various numerical techniques, and we will leave this as a direction to pursue for future work.

Given our primary focus on simple homogeneous liquids at ambient temperatures and densities, we introduce the Kirkwood superposition approximation[145] asserting that $\delta g_3(\mathbf{R}_1, \mathbf{R}_2, \mathbf{R}_3) = 1$ and therefore $g_3(\mathbf{R}_1, \mathbf{R}_2, \mathbf{R}_3) \approx g_2(\mathbf{R}_1, \mathbf{R}_2)g_2(\mathbf{R}_2, \mathbf{R}_3)g_2(\mathbf{R}_1, \mathbf{R}_3)$.[146] This approximation is known to hold at not too low temperatures and not too high enough densities, and it is conceivable that our liquid systems are suitable for employing this approximation, similar to previous work in CG systems of linear polymers.[147] Additionally, in line with Eq. (54), the homogeneous and isotropic nature of single-site CG models implies that the scalar variables $(R, S, T)$, or $(R, S, |R - S|)$, are spatial variables that can effectively represent the vectors of triplet configurations $(\mathbf{R}_1, \mathbf{R}_2, \mathbf{R}_3)$. Since the first term on the right-hand side of Eq. (53) vanishes, then $S_{\text{ex}}^{(3)}$ is reduced to

$$S_{\text{ex}}^{(3)} \approx \frac{8\pi^2\rho^2}{3} \int_0^\infty \int_0^\infty [g_2(R)g_2(S)g_2(|R-S|) - 3g_2(R)g_2(S) + 3g_2(R) - 1]R^2S^2 dRdS.$$

(55)

It is worth noting that due to the symmetry in Eq. (55), one practically needs to include all contributions from $(R, S, T)$ pairs as written in the brackets $[\cdot]$. Also, $T$ can be determined from $R$ and $S$. Numerically evaluating the two-dimensional integral $\iint dRdS$ can be done in a similar manner to Eq. (50) using Simpson's rule.

Yet, even with the existence of three-body corrections, higher-order terms beyond three-body effects are not accounted for in Eq. (55). Similar to Laird and Haymet's treatment of the pairwise entropy, we consider another correction factor proposed by Singh et al. to estimate the three-body CG entropy[109]

$$S_{\text{triplet}} = S^{(2)} + S^{(3)} - \frac{1}{3}\mathcal{R}.$$

(56)

Using Eq. (56), the evaluated $S^{(3)}$ and $S_{\text{triplet}}$ values for methanol and chloroform are listed in Table 7. It is immediately apparent that the three-body correction term $S^{(3)}$ is much smaller than the pairwise correction term $S^{(2)}$. The pairwise contribution takes up 78.74% of the overall entropy for methanol and 60.65% for chloroform, which aligns with trends reported in the literature.[83, 106-109, 148, 149] Namely, the three-body correction effect is more pronounced in chloroform than in methanol.

**Table 7.** Thermodynamic entropy for methanol and chloroform estimated using the multiparticle expansion approach in comparison with the reference entropy values obtained by employing 2PT calculations on actual CG trajectories. Here, the systematic expansion up to the third order, $S_{\text{triplet}}$, was considered with the additional correction term given by Eq. (56), while the Kirkwood superposition approximation was also introduced.

| Molecule | Thermodynamic Entropy Quantity (cal·mol$^{-1}$·K$^{-1}$) | | | |
| --- | --- | --- | --- | --- |
| | Actual CG Entropy | Ideal gas estimation | Three-body correction | $S_{\text{triplet}}$ |
| Methanol | 21.110 | 23.710 | -0.193 | 22.141 |
| Chloroform | 24.938 | 28.894 | -1.456 | 24.531 |

The performance of $S_{\text{triplet}}$ in methanol and chloroform reflects how the higher-order correction term compares to the pairwise contribution. In the case of methanol, which is dominated by pairwise correlations, $S_{\text{triplet}}$ is quite close to $S_{\text{pair}}$, differing by only 0.1 cal·mol$^{-1}$·K$^{-1}$. However, for chloroform, where higher-order correlations play a more significant role, the three-body correction greatly enhances the accuracy of CG entropy estimation. When considering only the pairwise correction for chloroform, there was an absolute error of 0.7-0.8 cal·mol$^{-1}$·K$^{-1}$, but this error was halved when accounting for the three-body correlations. Also, we would like to note that our estimation of $S_{\text{triplet}}$ is primarily based on the Kirkwood superposition approximation, and thus we anticipate further improvements when considering non-trivial three-body correlations. In summary, the systematic treatment based on multiparticle correlation functions offers an alternative to the hard sphere framework and the generalized van der Waals approaches. This systematic approach has the potential to enhance entropy evaluation in a practical manner by gradually including higher-order contributions.

## 4. Conclusions

Amidst the ongoing effort to better understand the role of entropy in CG modeling, our work is specifically aimed at developing an efficient framework for efficiently predicting CG entropy using only the FG statistics. This eliminates the necessity for CG simulations and additional sampling procedures. Drawing inspiration from classical perturbation theory of liquids, we designed two different approaches, as illustrated in Fig. 5.

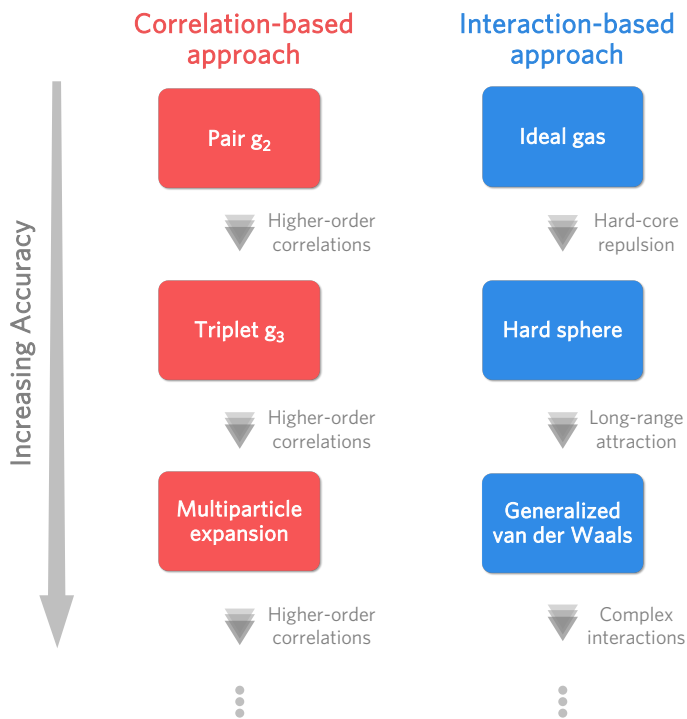

**Figure 5.** Overview of the systematic framework proposed in this work. In the pursuit of the accurate estimation of the entropy and other thermodynamic properties of CG models, two distinct approaches are possible. First, one can incorporate important interactions through a perturbative treatment of CG interactions (interaction-based approach). Alternatively, one can also decompose the target thermodynamic property via a many-body expansion and incorporate higher-order contributions to the model description (correlation-based approach).

For the interaction-based approach, starting from the simplest ideal gas approximation, we gradually incorporate important physics to correctly describe the entropy of non-ideal CG systems. This is achieved through a perturbative approach, ranging from adding hard-core repulsions, then adding generalized van der Waals interactions, and finally utilizing multiparticle expansion. Within each model, accuracy can also be enhanced by improving the model descriptions. For hard spheres, accurate EOSs would improve the fidelity of hard sphere models, and similarly, an accurate description of long-range attractive interactions or even complex interaction profiles using a multi-basin interaction form is expected to enhance the performance of the generalized van der Waals model. Alternatively, we can estimate the thermodynamic properties of CG models using the multiparticle expansion. In this approach, we can improve the model description by including higher-order contributions. A major advantage of this approach is that these structural correlations can be estimated from the atomistic reference, greatly reducing the computational overhead. In summary, we claim that achieving high fidelity predictions of thermodynamic properties can be accomplished by incorporating key physical descriptions of the system with accurate information.

Furthermore, while developing this framework, the perturbative treatment raises an important question: What are the features necessary to represent complex, many-body CG PMFs? Our approach addresses this problem by focusing on two fundamental yet simple interactions: hard sphere repulsion and square well attraction. For more complicated CG systems, the resultant CG PMF may contain other non-negligible interaction terms with complex profiles that differ strongly from that of a simple square well. Recently, we have demonstrated that many-body CG PMFs can be faithfully described as a sum of hard-core repulsion and Gaussian interactions by combining classical perturbation theory and integral equation theory. Therefore, one possible direction would be to extend the generalized van der Waals framework to CG models with Gaussian basis sets, which is expected to describe more complex energy landscapes of bottom-up CG models, including polymers and biomolecules.

Our findings provide valuable insights into the reliability of different physical approximations when estimating the entropy and other thermodynamic properties of molecular CG models. We have demonstrated here that the ideal gas description provides a reasonable prediction, while the hard sphere description performs better and yields accurate predictions for CG liquids. This is a noteworthy discovery considering the challenges posed by the non-trivial nature of CG interactions. Our proposed approach demonstrates that it is possible to construct an approximate CG partition function, incorporating the essential physics based on the FG statistics, through a bottom-up methodology. Notably, this framework's versatility allows for a straightforward extension to compute other thermodynamic properties. In summary, our work underscores the significance of adopting minimal representations of CG PMFs when predicting thermodynamic properties.

**ACKNOWLEDGMENTS**
J.J. thanks the Arnold and Mabel Beckman Foundation for financial support. J.J. also acknowledges insightful discussions on computing three-body correlations with Dr. Youngbae Jeon and thanks Dr. Ziwei He for useful comments.

**APPENDIX**
**A. Derivation of Eq. (31)**

We begin with the definition of the attractive coordination number, $N_c$, introduced in Eq. (30):

$$N_c = \frac{N}{V} \int_{attr} g(R) dR, \tag{57}$$

where $\int_{attr} dR$ integrates only the attractive part of the underlying potential. For example, in the case of a square-well potential, as defined in Eq. (29), $\int_{attr} dR$ corresponds to the volume integral

$$\int_{\sigma}^{R_\epsilon \sigma} dR = \frac{4\pi}{3} \sigma^3 (R_\epsilon^3 - 1), \tag{58}$$

due to spherical symmetry. Then, Eq. (58) can be simplified as

$$N_c = \frac{4\pi N}{3V} \sigma^3 (R_\epsilon^3 - 1) \frac{\int_{attr} g(R) dR}{\int_{attr} dR}. \tag{59}$$

Under the low-density condition, $\int_{attr} g(R) dR / \int_{attr} dR$ can be further simplified to be an effective Boltzmann weight applied to the average potential of mean force over the coordination volume, which can then be approximated as the PPMF ($e^{\beta \epsilon}$). Combining these equations, $N_c$ is reduced to

$$N_c = \frac{4\pi N}{3V} \sigma^3 (R_\epsilon^3 - 1) e^{\epsilon/kT}, \tag{60}$$

corresponding to Eq. (31).